\documentclass[conference,final]{IEEEtran}

\usepackage[hidelinks]{hyperref}
\usepackage{amsmath,amssymb}
\usepackage{cleveref}
\usepackage{xfrac}
\usepackage[binary-units=true]{siunitx}
\usepackage[nonumberlist,acronym,nomain]{glossaries}

\usepackage{todonotes}
\usepackage[utf8]{inputenc}

\usepackage[autostyle]{csquotes}
\makeatletter
\def\ps@IEEEtitlepagestyle{%
  \def\@oddfoot{\mycopyrightnotice}%
  \def\@evenfoot{}%
}
\def\mycopyrightnotice{%
{\footnotesize
  \begin{minipage}{\textwidth}
  \centering
  \textcopyright 2018 IEEE. Personal use of this material is permitted. Permission from IEEE must be obtained for all other uses, in any current or future media, including reprinting/republishing this material for advertising or promotional purposes, creating new collective works, for resale or redistribution to servers or lists, or reuse of any copyrighted component of this work in other works.
  \end{minipage}
  }
}

\newcommand{\eg}{e.\,g.\ }
\newcommand{\ie}{i.\,e.\ }
\newcommand{\cf}{cf.\ }

\makeglossaries
\newacronym{iot}{IoT}{Internet of Things}
\newacronym{itu}{ITU}{International Telecommunication Union}
\newacronym{iiot}{IIoT}{Industrial Internet of Things}
\newacronym{mc-ss}{MC-SS}{multi-carrier spread spectrum}
\newacronym{mccdma}{MC-CDMA}{multi-carrier code division multiple access}
\newacronym{mmtc}{mMTC}{massive machine type communications}
\newacronym{urllc}{uRLLC}{ultra-reliable and low latency communications}
\newacronym{embb}{eMBB}{extreme mobile broadband}
\newacronym{sdn}{SDN}{software-defined network}
\newacronym{rru}{RRU}{remote radio unit}
\newacronym{bs}{BS}{base station}
\newacronym{sgw}{S-Gw}{serving gateway}
\newacronym{pgw}{P-Gw}{packet data network gateway}
\newacronym{up}{UP}{user plane}
\newacronym{cp}{CP}{control plane}
\newacronym{ausf}{AUSF}{authentication server function}
\newacronym{udm}{UDM}{user data management}
\newacronym{hss}{HSS}{home subscriber server}
\newacronym{smf}{SMF}{session management function}
\newacronym{mme}{MME}{mobility management entity}
\newacronym{pcrf}{PCRF}{policy charging and rules function}
\newacronym{ap}{AP}{aggregation point}
\newacronym{prp}{PRP}{parallel redundancy protocol}
\newacronym{tcp}{TCP}{transmission control protocol}
\newacronym{api}{API}{application programming interface}
\newacronym{ovs}{OvS}{open vSwitch}
\newacronym{wlan}{WLAN}{wireless local area network}
\newacronym{wan}{WAN}{wide area network}
\newacronym{lpwan}{LPWAN}{low-power wide area network}
\newacronym{ble}{BLE}{Bluetooth low energy}
\newacronym{itur}{ITU-R}{International Telecommunication Union Radiocommunication Sector}
\newacronym{qos}{QoS}{quality of service}
\newacronym{nbiot}{NB-IoT}{narrow-band IoT}
\newacronym{ec}{EC}{edge cloud}
\newacronym{nfv}{NFV}{network function virtualization}
\newacronym{iira}{IIRA}{Industrial Internet Reference Architecture}
\newacronym{iic}{IIC}{Industrial Internet Consortium}
\newacronym{rami}{RAMI~4.0}{Reference Architectural Model of Industry 4.0}
\newacronym{mes}{MES}{manufacturing execution system}
\newacronym{erp}{ERP}{enterprise resource planning}
\newacronym{ngmn}{NGMN}{next generation mobile networks}
\newacronym{agv}{AGV}{automated guided vehicle}
\newacronym{agc}{AGC}{automated guided cart}
\newacronym{slam}{slam}{simultaneous locating and mapping}
\newacronym{mno}{mno}{mobile network operator}
\newacronym{plc}{PLC}{programmable logic controller}
\newacronym{epc}{EPC}{evolved packet core}
\newacronym{indot}{ind. OT}{industrial operation technology}

\begin{document}
%
\title{A 5G Architecture for The Factory of the Future}

\author{\IEEEauthorblockN{Stephan Ludwig\IEEEauthorrefmark{1},
Michael Karrenbauer\IEEEauthorrefmark{2},
Amina Fellan\IEEEauthorrefmark{2},
Hans D. Schotten\IEEEauthorrefmark{2},
Henning Buhr\IEEEauthorrefmark{3},\\
Savita Seetaraman\IEEEauthorrefmark{3},
Norbert Niebert\IEEEauthorrefmark{3},
Anne Bernardy\IEEEauthorrefmark{4},
Vasco Seelmann\IEEEauthorrefmark{4},
Volker Stich\IEEEauthorrefmark{4},
Andreas Hoell\IEEEauthorrefmark{5},\\
Christian Stimming\IEEEauthorrefmark{5},
Huanzhuo Wu\IEEEauthorrefmark{6},
Simon Wunderlich\IEEEauthorrefmark{6},
Maroua Taghouti\IEEEauthorrefmark{6},
Frank Fitzek\IEEEauthorrefmark{6},
Christoph Pallasch\IEEEauthorrefmark{7},\\
Nicolai Hoffmann\IEEEauthorrefmark{7},
Werner Herfs\IEEEauthorrefmark{7},
Elena Eberhardt\IEEEauthorrefmark{8} and
Thomas Schildknecht\IEEEauthorrefmark{8}}\\

\IEEEauthorblockA{\IEEEauthorrefmark{1}Robert Bosch GmbH, Corporate Sector Research and Advance Engineering,\\70465 Stuttgart, Germany, stephan.ludwig2@de.bosch.com}

\IEEEauthorblockA{\IEEEauthorrefmark{2}Chair for Wireless Communications and Navigation,\\University of Kaiserslautern, Germany, \{karrenbauer,fellan,schotten\}@eit.uni-kl.de}

\IEEEauthorblockA{\IEEEauthorrefmark{3}Ericsson GmbH, Herzogenrath, Germany,
\{henning.buhr,savita.seetaraman,norbert.niebert\}@ericsson.com}  

\IEEEauthorblockA{\IEEEauthorrefmark{4}FIR e.V. at RWTH Aachen, Germany,
\{anne.bernardy,vasco.seelmann,volker.stich\}@fir.rwth-aachen.de}

\IEEEauthorblockA{\IEEEauthorrefmark{5}SICK AG, Waldkirch, Germany,
\{andreas.hoell,christian.stimming\}@sick.de}

\IEEEauthorblockA{\IEEEauthorrefmark{6}Deutsche Telekom Chair of Communication Networks,\\
University of Dresden, Germany, \{huanzhuo.wu,simon.wunderlich,maroua.tahouti,frank.fitzek\}@tu-dresden.de}

\IEEEauthorblockA{\IEEEauthorrefmark{7}Laboratory for Machine Tools and Production Engineering at RWTH Aachen, Germany\\
\{c.pallasch,n.hoffmann,w.herfs\}@wzl.rwth-aachen.de}

\IEEEauthorblockA{\IEEEauthorrefmark{8}Schildknecht AG, Murr, Germany, 
\{elena.eberhardt,thomas.schildknecht\}@schildknecht.ag}
}



\maketitle

\begin{abstract}
Factory automation and production are currently undergoing massive changes, and 5G is considered being a key enabler.
In this paper, we state uses cases for using 5G in the factory of the future, which are motivated by actual needs of the industry partners of the \enquote{5Gang} consortium.
Based on these use cases and the ones by 3GPP, a 5G system architecture for the factory of the future is proposed.
It is set in relation to existing architectural frameworks.
\end{abstract}


%
\IEEEpeerreviewmaketitle

\section{Background}
The research initiative \enquote{5G: Industrial Internet} \cite{bundesministerium_fur_forschung_und_bildung_5g:_2018} of the German Federal Ministry of Education and Research, addresses the requirements imposed on a 5G communication network in order to be used with applications of industrial production.
As part of this initiative, the project \enquote{5Gang} (\url{http://5gangprojekt.com}) considers different use cases of future industrial production principles and their requirements on the communication network. 
\enquote{5Gang} is a consortium of eight partners from industry and academia who bring in experience covering not only technical but also business and production process aspects.
The project scope does not only cover local production sites but also the opportunities arising from adding inter-site connections. 
For two use cases demonstrator platforms will be built in a later project phase in order to validate and optimize their performance in a 5G mobile network. 

\section{Introduction---5G in Production}
Factory automation and production are currently undergoing massive changes, which is termed the $4^\text{th}$ revolution after the mechanization using water and steam ($1^\text{st}$), the introduction of mass production using the conveyor belt and electric energy ($2^\text{nd}$) and the digitalization with electronics, robots and information technology ($3^\text{rd}$).
This change is also termed Industry 4.0 
or \gls{iiot} and it is characterized \eg by connected machinery, sensors and humans within an \gls{iiot} as well as by decentral decision taking using cyber physical systems and a merge of real and virtual worlds.
The vision is to allow individualized products (lot size 1) at the cost of that of mass production (\enquote{mass customization}).
For classic factories doing mass production, their production costs will be significantly reduced by following along the path towards Industry 4.0.
This requires a reduction in production time, an increase in automation and a highly flexible and reconfigurable kind of production.

The connectivity of production devices has to become likewise flexible.
This is why the $5^\text{th}$ generation of mobile, cellular networks (5G) turns out to become a key enabler for new use cases towards the vision of Industry 4.0.
The full standardization of 5G is planned to be finished by 2020, but commercial pre-standard systems will be deployed already in 2018.
Release 15 of 3GPP, which was frozen by the end of 2017, is considered as a pre-version of 5G.
Requirements might change on its way, but the overall targets are taking shape and can be considered as a base for \enquote{5Gang}.

Performance is one major aspect when analyzing the possibilities to use 5G in a production environment.
The United Nations organization \gls{itur} addresses three main capabilities to define the requirements for 5G systems \cite{itu-r_study_group_05_working_party_5d_itu-r_2017}:
\begin{itemize}
  \item \Glsfirst{embb}\glsunset{embb} will provide up to \SI{20}{\giga\bit\per\s} of data from the end users towards the network. 
  The user plane latency shall be below \SI{4}{\ms}.
  \item \Glsfirst{urllc}\glsunset{urllc}, or critical machine type communication, requires a user plane latency of less than \SI{1}{\ms}.
  \item \Glsfirst{mmtc}\glsunset{mmtc}, will allow to connect up to one million devices per \si{km^2} with the given \gls{qos}.
\end{itemize}
To some extend, other existing wireless technologies can fulfill the requirements 1 and 3 already today. 
However, they cannot reliably support the \gls{urllc} case.

Apart from the new radio interface, 5G will impose massive changes by utilizing softwarization and virtualization of and in the core network.
However, this change will not come abruptly, but 4G networks will continuously transit to these technologies.
Some further advantages of 5G networks (not complete) are planned to be:
\begin{itemize}
  \item Convergence: The same physical network can be used for many use cases.
  \item Support for mobility: Moving workpiece carriers can not only be controlled or tracked inside the factory but also on their way between different production sites.
  \item Long battery life: 
  Allow operating times of 10 years for transmitting small volumes of data from battery powered devices in an energy efficient manner.
  \item Privacy: It will be possible to install mobile network equipment in a cost-efficient manner inside the factory.
  This will prevent the risk of production parameters or other data leaving the factory site.
  \item Security: SIM cards (or similar solutions) provide a secure way to manage devices and restrict network access.
  \item Economy of scale: The big 5G ecosystem will increase the volume of the radio modems leading to cheaper equipment, which would not be possible if every production solution used separately specialized hardware.
\end{itemize}

In the core network, some of the required key technologies are already available today, some others will be developed step by step.
One is network slicing \cite{3gpp_architecture_2014,3gpp_enhancements_2016}, which lets one reserve resources per use case and provides the required \gls{qos}.
Another one is distributed cloud or \gls{ec} which allows to move certain functions closer to the end-user using \gls{nfv}.
In a production environment this can even mean to deploy not only base stations or \gls{rru} in the factory but also core network functions.
This would not be reasonable with specialized and high-performance mobile nodes, but the latest virtualization techniques can bring down the costs.
Therefore, 3GPP has as well standardized the option to move only the user plane closer to the network edge (\cf \cite{3gpp_architecture_2018}).
Up to now, no 5G system architecture in the industrial context has been proposed.
Only an architecture for the general 5G cellular system was developed by the \enquote{5G Architecture White Paper} \cite{5gppp_architecture_working_group_view_2017} by the EU project \enquote{5GPPP}.
This paper proposes such a general and flexible system architecture for a 5G network in the factories of the future.
The rest of this paper is structured as follows:
\Cref{sec:use_case} summarizes the use cases considered in \enquote{5Gang} and maps them onto the use cases found by 3GPP.
In \cref{sec:frameworks} we describe how the architecture integrates into existing frameworks.
Afterwards, in \cref{sec:arch} the 5G architecture is proposed and its components and functions are described.

\section{Use Cases for 5G in Industry 4.0}\label{sec:use_case}
In the \enquote{5Gang} project, use cases were researched based on the wide spectrum of actual needs stated by the project partners.
The use cases were grouped into use case classes in order to define a framework for systems with similar requirements.
An overview of these use case classes is given in this section.

\subsection{Smart Production}
Due to the transition to flexible manufacturing and customization of the production processes, customers increasingly ask for a dynamic order-change request system along the whole supply chain.
This demands reliable tracking of the desired product, where orders and materials have to be tracked across different locations of the supply chain located all across the globe.
Such tracking is enabled by a battery-powered wireless sensor device mounted on a workpiece carrier.
5G is in this case the enabling technology, because it supports the low-energy needs of battery powered devices and it will connectivity all over the globe.
In a second use case a manufacturer aims to detect defects in parts at even earlier stages through inline quality control.
Avoiding unnecessary processing of defect parts can save further costs when being avoided.
A similar case occurs when tracking items within a logistics chain.
Here, the transport conditions like detecting shock events or monitoring the cooling temperature in the case of food are monitored continuously.
In the event of significant deviations, a re-order can be made before the items arrive and their transport damage is detected.
In both cases, sensors, placed on the workpiece or on the workpiece carrier, transmit their measured values to a central point for evaluation.
A fourth use case is a shared tracking and tracing system, which stocks material in a shared warehouse.
So far, the delivery of materials in a company requires a manual check-in followed by the search for a suitable stocking place.
For all four use cases in this class, the massive number of connected sensor devices will be handled in 5G by \gls{mmtc}.

\subsection{Automated Guided Driving}
The challenge in a flexible production environment is to quickly adapt to changes in production conditions or environment in order to be able to react to unexpected system behavior in real-time.
This class includes applications with different \glspl{agv} transporting material in a dynamic environment like with \glspl{agc}.
The routes of \glspl{agc} can be variable in general. In a well-planned production, however, the material flow follows a milk run principle with dynamic changes of the milk run route.
In some cases, \glspl{agc} can be controlled in a master-slave relation, \ie only the master \gls{agc} needs to communicate with the production system, while the slave \glspl{agc} are connected to the master in their own network.
Another use case has high requirements on the available bandwidth, in order to allow reliable navigation of \glspl{agv} or mobile production robots in a spatially varying production environment using high-precision real-time maps.
\Glspl{agv} can cooperate to simultaneously localize themselves and jointly create the maps, termed \gls{slam}, which can even be in a distributed fashion.
Each \gls{agv} provides its own measurements to other devices in order to create the common map.
The measurements should be timely synchronized and shared among the devices, so that they can simultaneously locate themselves and plan further movements.
Therefore, a highly reliable transmission of massive data at low latencies is indispensable.
5G is an enabler for this scenario because it offers to reserve parts of the network with a guaranteed \gls{qos}.

\subsection{Distributed Sensing and Condition Monitoring}
In production, unexpected machine defects cause downtimes, which result in high costs and delay delivery.
Condition Monitoring refers to the supervision of machine conditions (the state of the factory) in order to prevent unplanned defects, to detect abnormal behavior and to be able to plan maintenance activities.
Furthermore, the quality of products depends on the condition of machines which produce them.
If the tools inside a machine wear down, the quality of the product often decreases without a visible reason.
Thus, recording condition characteristics along the production process enables a continuous quality control of the products.
To this, a large number of small, wireless sensor nodes (\eg microphones, wearable sensors, cameras, temperature sensors) are distributed in the plant.
They can be allocated to dynamic processes and are moderately mobile, as they can occasionally change their positions, \eg during reconfigurations or rebuilds.
The sensors are required to have a long battery life time, since they can be placed in hardly accessible locations.
The massive number of wireless sensor nodes will require an interference management as with the 5G \gls{mmtc} profile as an enabler.
In some use-cases highly mobile broadband connectivity might be required, \eg if a high resolution cameras need to be integrated into a machine on a moving part.

\subsection{Infrastructure Retrofit}
The benefits of 5G for the infrastructure retrofit refers to three sub problems. 
Existing sensor or actuator technologies can be enhanced by the integration of a mobile communication technology module in any of the three flavors of 5G, \gls{embb}, \gls{urllc} and \gls{mmtc}.
Furthermore, in use cases like the decentralized measurement of gas-flow through pipelines, 5G can provide the connections from the central monitoring and control to the measurement spot and decentralized equipment, which have long distances between each other.

\section{Integration into and Relations to Existing Frameworks}\label{sec:frameworks}
For the vertical integration of production technologies, the two most-cited reference architectures for the \gls{iiot} are the \gls{iira} of the \gls{iic} and the \gls{rami}, into which we integrate our proposed architecture in this section.
Furthermore, we show relations to the \enquote{5G Architecture} of 5GPPP and the considerations of 3GPP.

\subsection{NGMN 5G White Paper}


One of the first white papers that brought up 5G for a vertical was the \enquote{5G White Paper} \cite{ngmn_alliance_5g_2015} of the \gls{ngmn} consortium.
It addresses verticals and envisions the use cases of massive sensor networks in the \gls{iot}, which is termed \enquote{massive IoT}; of ultra-reliable communication; and of extreme real-time communication, \eg for collaborative robots.
All three use cases are also considered in \enquote{5Gang} and require security, identity and privacy; real-time, seamless and personalized experience; responsive interaction and charging as well as \gls{qos} and contextual behavior or the system.
Beyond connectivity, a 5G system could offer services like transparency of connectivity, location, resilience, reliability and high availability.
Our proposed architecture integrates and details the (virtualization of the) infrastructure resource layer and parts of the business enablement layer of the \enquote{5G architecture} in \cite[sec.~5.4]{ngmn_alliance_5g_2015}

\subsection{IIRA}

\begin{figure}
  \centering
  \includegraphics[width=0.8\columnwidth]{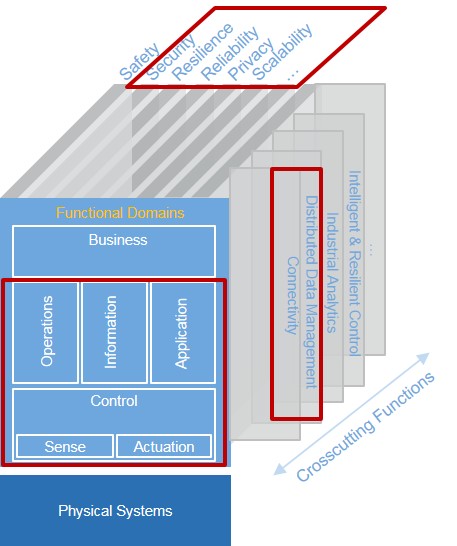}
  \caption{\glsentrylong{iira} \cite[Fig.~6-5]{_systems_2011}}
  \label{fig:iira}
\end{figure}

The \glsentryfull{iira} \cite{lin_industrial_2017} of the \gls{iic} follows ISO/IEC/IEEE 42010:2011 \enquote{Systems and Software Engineering – Architecture Description} \cite{_systems_2011} and contains \gls{iiot} architecture framework, which in turn contains views on stakeholders, concerns, viewpoints, and model.
In general, the \gls{iira} consists of three dimensions: functional domains, system characteristics and crosscutting functions \cref{fig:iira}. 
It considers functional domains, namely control, operations, application, business and information.
As highlighted in \cref{fig:iira}, our proposed architecture covers all functional domains except the business domain and all system characteristics except safety.
Our architecture addresses especially the information functional domain as 5G allows efficient and high-performance communications. 
%
%
The 5G system allows information gathering and is part of the cross-cutting functions, however, limited to the functions connectivity and distributed data management contained therein.
%


\subsection{RAMI4.0}
\begin{figure}
  \centering
  \includegraphics[width=\columnwidth]{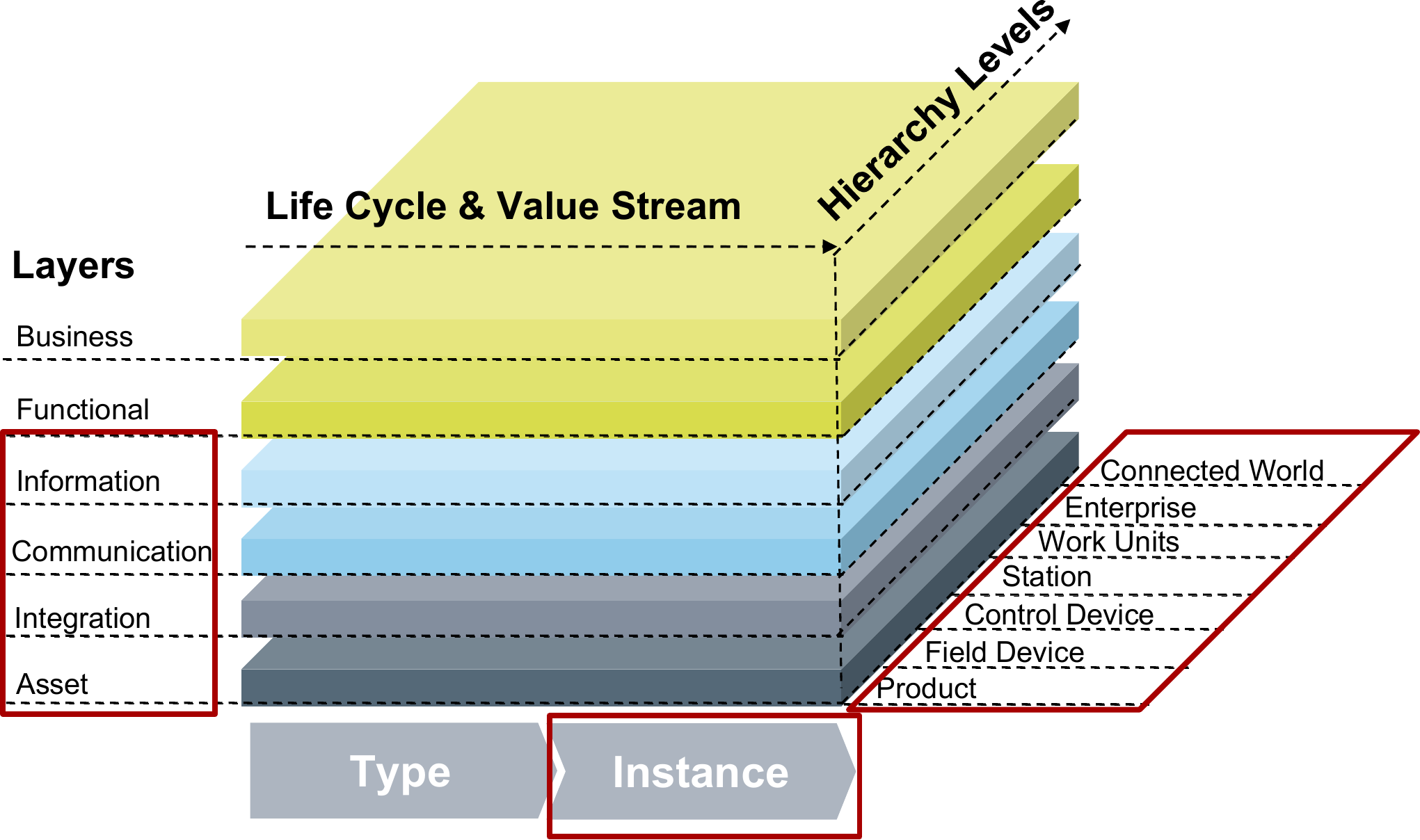}
  \caption{\glsentryfull{rami}}
  \label{fig:rami}
\end{figure}

The \enquote{\glsentrylong{rami}} (\glsentryshort{rami}) \cite{martin_reference_2015,adolphs_rami_2015} was designed as a future reference model for industrial production and automation to categorize and differentiate different architectural views that are related to each other.
The \gls{rami} is structured as a three dimensional model comprised of the axes hierarchy levels, layers and life cycle value stream \cref{fig:rami}.
The hierarchy levels represent the classic automation pyramid, which structures different layers of responsibility and aggregation from field devices over control hardware to higher-level applications, \eg \gls{mes} and \gls{erp}.
The hierarchy levels enhance the classic pyramid by the product itself and the connected world including the possibility of smart (\ie communicating) products as well as connecting enterprises or shop floor software technologies to cloud technologies.
The industrial networking architecture presented in this paper covers all hierarchy levels \ie its offered functionality addresses every layer in the automation pyramid and the connection to super-ordinate cloud services.

The layers axis divide the solution into six functional levels.
The asset layer addresses all physical objects available at production sites including field devices, work pieces and workers.
The integration layer covers all technological methodologies for digitally integrating assets (\eg attaching QR-Codes onto work pieces or using RFID-ID). 
The communication layer describes the way which communication technology is used for exchanging or accessing data for underlying assets. 
The information layer defines how data is represented and the functional layer describes functionalities for different assets or general system functions. 
Finally, the business layer specifies business related information exchange as well as business processes for given production or process segments (\eg engagement times, down-times, jobs and orders, amount of produced goods). 
Our proposed architecture covers the layers from the asset to the information layer. 

The life cycle value stream (of products) axis divides the product development and usage process into a type and an instance phase. 
Whereas the type phase refers mainly to product development including documentation, construction plans etc., the instance phase refers to the usage phase of the product, in which data is collected during its operational phase. 
The architecture presented in this paper covers the instance phase of a product life-cycle, which is more challenging from a communication point of view.

\subsection{5G Architecture White Paper of 5GPPP}
\begin{figure}
  \centering
  \includegraphics[width=\columnwidth]{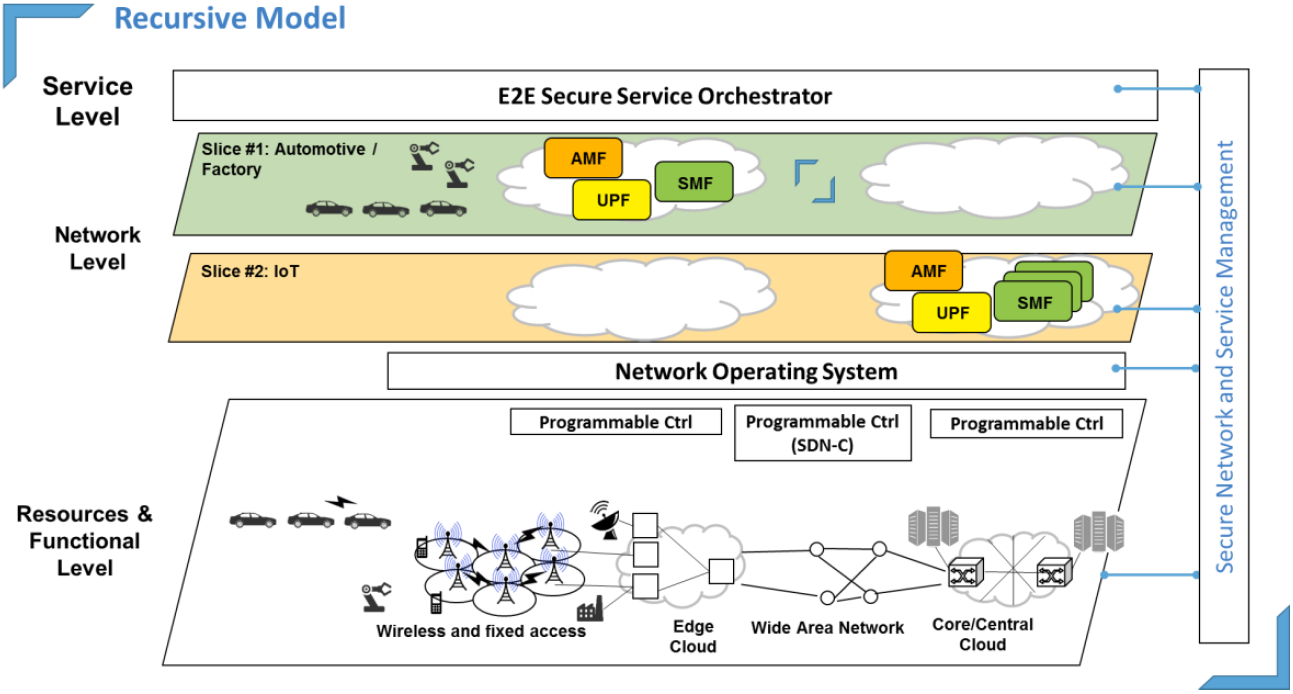}
  \caption{5GPPP Overall Architecture \cite[Fig.~2-1]{5gppp_architecture_working_group_view_2017}}
  \label{fig:5gppp}
\end{figure}

An architecture for 5G was developed by the EU-funded project \enquote{5GPPP}  in \cite{5gppp_architecture_working_group_view_2017}.
It primarily focuses on network slicing and the radio access.
Furthermore, it gives a detailed overview of the capabilities of each component and describes the infrastructure \cref{fig:5gppp}, into which the \enquote{5Gang} architecture integrates for use in industrial applications and in the factory of the future.

The overall architecture is divided into resource/functional level, network level and service level.
The resource \& functional level provides the physical resources for communication, computation and storage from the network access to the core network and Internet.
Apart from the wireless and fixed access it consists of edge cloud and core/central cloud resources.
Towards the network level, virtualization of the physical network is achieved by a network operating system and programmable network control units as with \gls{sdn} such that network slices can be build on top.
On the service levels these slices are orchestrated in an end-to-end fashion using  management functions of all levels.
In order to integrate into this overall architecture, our 5G system enables network slicing by providing the physical network infrastructure and corresponding management and orchestration interfaces.

\subsection{Requirements from 3GPP TR 22.804}
In preparation of Release 16, 3GPP performed a study on communication for automation in Vertical Domains.
The resulting report \cite{3gpp_study_2017} names use cases and their requirements for the 5G system of 3GPP.
Apart from other verticals, one major part is on the use case class \enquote{factories of the future} and industrial security requirements.
Within these, \cite{3gpp_study_2017} addresses similar use cases as \enquote{5Gang}.
Motion control, and mobile robots are part of the use case class \gls{agv}.
The 3GPP use cases massive wireless sensor network and process automation match with the distributed sensing and control class.
Remote access, plug and produce for field devices, flexible/modular assembly area, control-to-control communication and inbound logistics for manufacturing, parallel with the class smart production, while connectivity on the factory floor overlaps with infrastructure retrofit.
Use cases like motion control, mobile control panel, augmented reality and process automation/closed-loop control were not described in this paper as they are addressed by other projects in the research initiative.
However, their requirements were incorporated into the designed architecture.
In contrast, (quality) tracking along the production line and large area network as with the pipeline use case were not considered so far by \cite{3gpp_study_2017}.
For the overlapping use cases, it was found that the requirements of \enquote{5Gang} and \cite{3gpp_study_2017} are well-aligned such that this paper can be considered to detail the first approach of the principle architecture of the \cite{3gpp_study_2017}.

\subsection{3GPP Architecture}
3GPP also proposed a first architecture of its 5G system in \cite{kim_3gpp_2017}.
It details the 5G components on the network side, \eg base stations and core network, assuming that 5G-enabled devices will have an anyhow natured 5G modem.
In this paper, we embed the architecture of \cite{kim_3gpp_2017} into the factory of the future and abstract it as the \enquote{cellular backend} (\cf \cref{fig:cell_backend} and next section).

\section{5G System Architecture}\label{sec:arch}
\begin{figure}
  \centering
    \includegraphics[width=0.8\columnwidth]{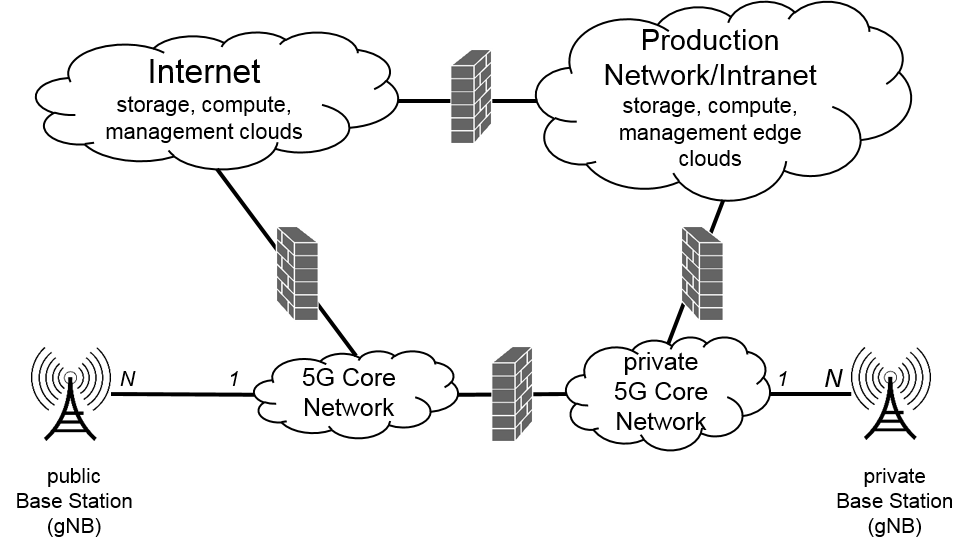}
  \caption{5G Architecture Overview of the \enquote{cellular backend}}
  \label{fig:cell_backend}
\end{figure}

In this section, we propose the 5G architecture for the factory of the future, which addresses the requirements of the named use case classes.
A component view of the architecture is shown and a functional description is given.
The \enquote{cellular backend} of \cite{kim_3gpp_2017} is abstracted as shown in \cref{fig:cell_backend}, which contains two main parts, the classic public cellular network run by an \gls{mno} and a private cellular network.
While the public network can be installed on-site or outside the production facilities, the private network is typically installed on-site only.
It can be operated by the factory owner or a third-party, including classic \glspl{mno}.
Its main feature is that it directly connects to the company IT infrastructure without the data traveling through any public network.

In both cases, for a public and a private network, the user equipment connects to base stations, where the term base stations stands for either a \gls{bs} or baseband unit plus \gls{rru}, where multiple \glspl{rru}/antennas can connect to one base station via cable or wireless backhaul.
The base station contains a firewall and L2/L3 routers (implementing a cloud-\gls{sdn}), \gls{sdn} controller and resource management for non-5G resources/links.
Multiple base stations are interconnected through the core network \cite{kim_3gpp_2017}, which also offers classic services like \gls{smf}, \gls{pcrf}, or \gls{ausf} and \gls{udm}---the functions of the former \gls{hss} (LTE)---as well as access and mobility management---the former \gls{mme}.
Furthermore, the 5G core network can provide (edge) cloud services, \ie storage, compute and management of devices and the functions of the core network can be virtualized such that they can run on any (edge) cloud.
The \gls{epc} contains \gls{sgw} and \gls{pgw} might have a split of user and data plane.
In a public 5G core network, the data plane typically interfaces the Internet for further services, while in the private 5G core it can be broken out to the production network or the company network, which then again might interface the public Internet.
Such services could be a device cloud, which handles tasks like user, device, and access rights management; setting parameters in the connected devices; providing user interfaces in order to further process data in \gls{erp} systems or other data cloud services by, \eg, using a RESTful \gls{api}.
Further functions are alerting, sending and receiving confirmation via SMS, field strength display of the current data connection or creation of documentation and their delivery to authorized recipients.
Above that, the private network can offer a interface to a public 5G core in order to handle other traffic, \eg from third-party contractors.
Of course, all connection between networks are guarded by firewalls, which allow to embed rule on which kind of data may cross interfaces.

\begin{figure}
  \centering
    \includegraphics[width=\columnwidth]{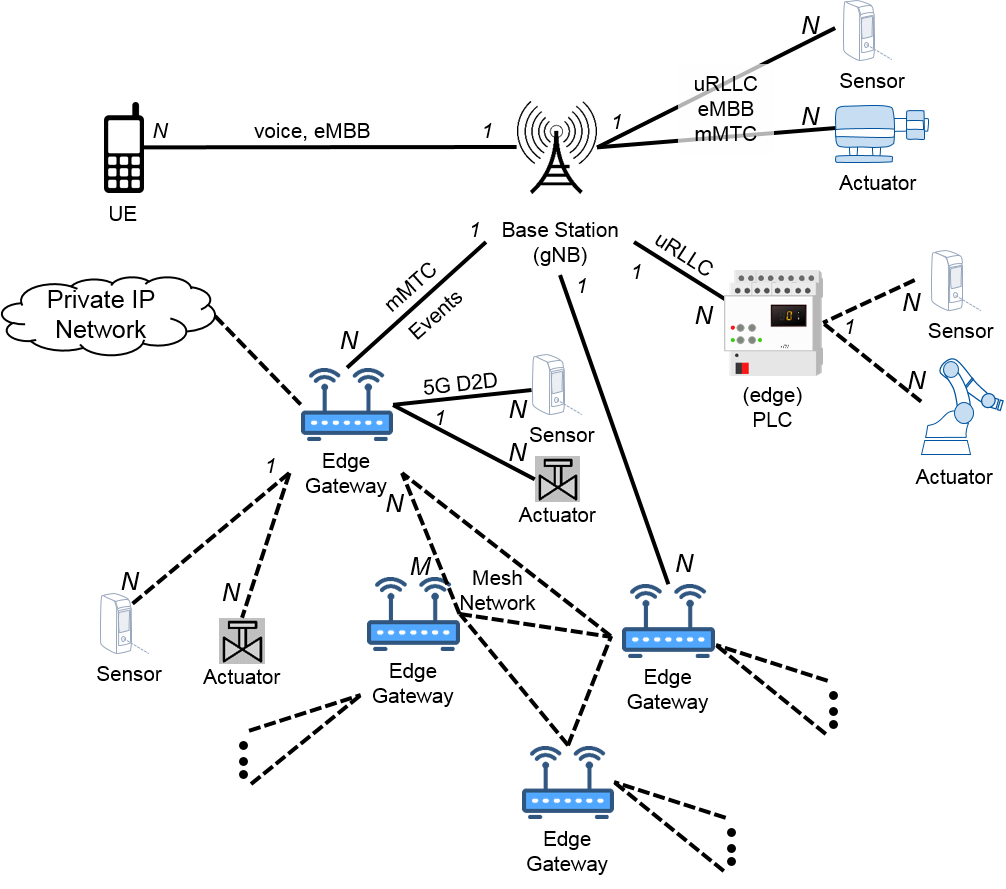}
  \caption{5G Architecture Overview}
  \label{fig:arch_layer1}
\end{figure}

The 5G architecture of the actual factory of the future is shown in \cref{fig:arch_layer1}, where different types of user equipment connects to public or private base stations: \gls{iiot} edge gateways, (edge) \glspl{plc}, actuators and automation sensors as well as wireless sensor devices, which will be detailed in the sequel.
In the figure, 5G connections are denoted by straight lines, while other technology links are indicated as dashed lines.
In the architecture, there is also classic user equipment like mobile phones for voice or \gls{embb} connections, which use 5G as replacement for any other cellular technology.
Then, there are general actuators and sensors, which are directly connected to the base station.
They directly address the use cases of \cite{3gpp_study_2017}, \eg closed-loop motion control using \gls{urllc}, augmented reality applications via \gls{embb} or massive wireless sensor-actuator networks using \gls{mmtc}.
In this case the control software runs in the edge cloud, which might be provided by the base station, as cloud-\gls{plc}.

Further devices can be connected to the base station.
(Edge) \glspl{plc} and edge gateways in turn provide connectivity to subsequent devices.
An edge gateway can use two modes depending on the system requirements: data transmission of sensor data with \gls{mmtc} technology or just sending event/update messages.
Edge gateways may be interconnected with each other and can form a mesh network such that the network reliability and their range can be extended as not every gateway requires 5G access then.
In contrast, (edge) \glspl{plc} can use the 5G-\gls{urllc} link for inter-\gls{plc} communication, \eg within a manufacturing cell.

\subsection{(Edge) Programmable Logic Controller}
A \gls{plc} is a real-time capable processing unit running as dedicated hardware or as a real-time core on an industrial computer, which is used as central logic for controlling automation processes.
They are designed to handle a variety of input and output interfaces in form of electrical analog and digital signals (mainly \SI{24}{\V}) as well as serial and Ethernet-based communication interfaces.
Industrial automation sensors and actuators, referred to as field devices, are signal data generators and consumers that are directly connected using electrical signals or fieldbuses, thus forming a bus network topology.
The main task of a \gls{plc} is to acquire data from field devices, apply logic and arithmetic functions and set output values based on the computations.
In order to assure correct and valid execution as well as short production cycles, automation steps, especially those involving linear or rotary movements, need to be continuously monitored and executed within short time frames, which typically requires round-trip times in the range of several milli- to microseconds.

\begin{figure}
  \centering
  \includegraphics[width=0.6\columnwidth]{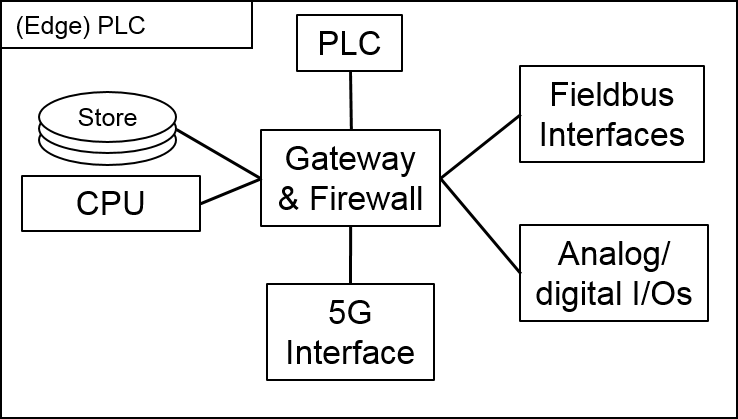}
  \caption{Edge \glsentryshort{plc} Component View}
  \label{fig:edge_plc}
\end{figure}

However, as \glspl{plc} were designed for primarily interfacing field devices, an integration into higher-level cloud-based services was originally not intended.
Furthermore, accessing production processes, automation cells or machines in order to adapt to production parameters, exchange industrial control software or update control algorithms is unfeasible or associated with high effort with existing control paradigms.
Hence, new approaches combining industrial automation with cloud-based services are necessary to gain benefits.
Using an intermediate edge layer enables leveraging the computational power of the cloud down to the shop floor.
This requires a new paradigm shift splitting the industrial control environment into industrial information technology (industrial IT, also cloud-\gls{plc}) and \gls{indot}.
Whereas, industrial IT resides mainly as software components and modules, which can be deployed and executed on different systems running locally or in the edge cloud, \gls{indot} involves industrial hardware components as well as some store and compute resources (edge \gls{plc}, \cf \cref{fig:edge_plc}).

Apart from compute and store resources, it contains the components of a classic \gls{indot} hardware, \ie \gls{plc} logic processor, wired fieldbus interfaces and analog/digital I/Os.
Furthermore, a 5G modem is available for uplink communication to other \glspl{plc} (inter-\gls{plc} communication) or the industrial IT.
All components are inter-connected through a gateway (maybe a \gls{sdn} switch, \cf \cref{sec_sdn_switch}) and a firewall, which enforces the data policy.

\subsection{Automation Sensors}
Electrical sensors are used in various industrial applications for detection, measurement, identification, inspection, etc.
In general the available sensors for an automation environment can be divided into three groups:
The simplest type of sensors uses \SI{1}{\bit} digital I/O capabilities (on vs.\ off).
Both of them have to be initialized with separate thresholds for switching.
Therefore they need maintenance or calibration at setup as well as in continuous operation.
In contrast, analog sensors report continuous levels of information, most commonly potential levels between \SI{0}{\V} and \SI{24}{\V}.
Thirdly, more advanced automation sensors use industrial fieldbus protocols to communicate with other members of their bus system.
They can offer simple \glspl{api} for data acquisition from master \glspl{plc}. If the bus master does not need to know the exact sensor's value, pre-processing can be done to allow fine-grained control over the process and just report special events over the fieldbus.
State of the art sensors cannot be used by multiple users without modification leading to additional circuitry in the electric/control cabinet or in the sensor itself.
In addition to that, if both cells have a different reference voltage levels, they cannot interoperate.

\subsection{Actuators}
Actuators can be found throughout an automated production facility to transform electrical energy and signals to motion of mechanics, fluids or other mediums, \eg in robotic applications, conveyor belts or grippers.
The actual robot consists of multiple motor-controlled axes.
The actual movement could be realized by stepper or DC-motors, which can be operated in an open or closed-loop scenario.
The former does not consist of a loop-back and relies only on the estimated position of the motion controller, while the latter uses encoder signals to retrieve the actual position.
Stalling or lost steps can only be detected by closed-loop control.
By the means of an additional \gls{plc}, which is connected over a fieldbus or industrial Ethernet in most cases, the motion controller receives new movement commands leading to the desired motion path.

\subsection{Wireless Sensor Devices}
Wireless sensor devices combine a wireless data transmission technology (connectivity) and a set of typically either sensors or actuators. 
They are typically powered by an on-board battery in order to be mountable almost everywhere. 
Because of the massive amount of devices as, \eg, in a production hall with the condition monitoring use-case, the battery power must last for several years without re-charge. 
Typical sensors are linear acceleration, rotational speed, and geomagnetic field orientation, each in 3 axes, where these sensor data are fused into an agile orientation value.
Further sensor data can be light intensity, acoustic (noise)/microphone, humidity, temperature, GPS position, smoke or specific gases.


\subsection{IIoT Edge Gateway}
\begin{figure}
  \centering
    \includegraphics[width=0.6\columnwidth]{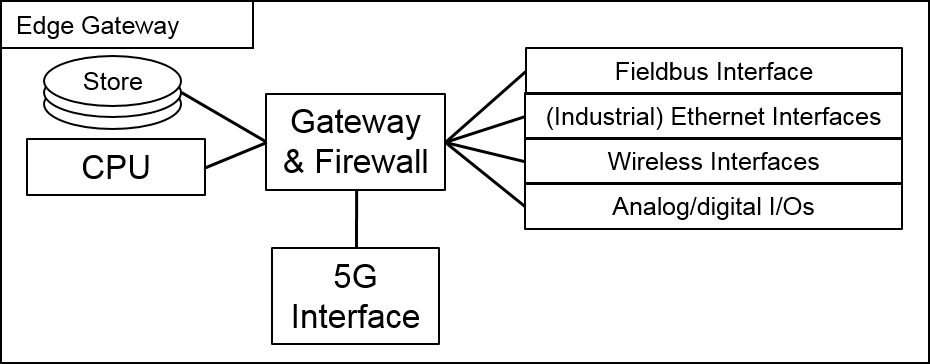}
  \caption{\glsentryshort{iiot} Edge Gateway Component View}
  \label{fig:edge_gw}
\end{figure}

The \gls{iiot} edge gateway, or \gls{ap}, primarily provides connectivity to networks with a large number of sensors, controllers and actuators, all of which could be using different wireless/wired technologies.
It is thus located at the edge of the 5G network near the data provider and at the edge of the network leading from the machine to the cloud.
It interfaces sensors and actuators and provides them with further connectivity, \eg to a (edge) cloud service or a service within the production environment/company network.
Therefor it contains a 5G modem and includes different wireless as well as wired interfaces (\cf \cref{fig:edge_gw}) like analog/digital I/Os (\eg classic \SIrange{4}{20}{\mA}/\SIrange{0}{10}{\V}), (Industrial) Ethernet (\eg PROFIBUS), Ethernet-based fieldbuses (PROFINET, Sercos III, CAN, Modbus, etc.), wireless (Bluetooth low energy, IEEE 802.11 WiFi). 
As a special case, a 5G device-to-device link is used, where sensors and actuators share the 5G connection of the gateway.
Furthermore, the gateway contains storage and processing resources, which might offer edge cloud services to the connected devices.
Alternatively, these resources can be used to reduce the available amount of data and, thereby, to optimize the transport by the network in terms of speed and costs. 

From a production point of view, the \gls{ap} shall interface to the process, \ie collect the operating and diagnostic data of the machine or its sensors together with a time stamp.
The compute and storage resources provide maximum flexibility to the user:
The data can be cached, pre-processed and analyzed on-size such that, \eg, only alarms are forwarded through the 5G interface.
In addition, the type of transmitted data and the size of packets can be dynamically managed and different modes of link operation can be used in order to adjust connectivity costs: an online mode with permanent connection; an interval mode which transmits only regularly at specified periods; and a sleep mode where the device transmits only if needed.
Furthermore, classic IT security mechanisms can be implemented using transport layer and as end-to-end encryption encryption, even by . 
As special cases, the processed data can be sent through the mobile radio provider with the strongest signal (\enquote{unsteered roaming}).

The \gls{ap} employs \gls{sdn} solutions (\cf \cref{sec_sdn_ctrl,sec_sdn_switch}) in order to manage the traffic within and between its sub-networks of connected devices.
It comprises of a \gls{sdn}-capable switch (\eg Open vSwitch), which facilitates traffic management, and an \gls{sdn} controller, which holds a centralized logical view of its network. 
A radio management system entity manages the large number of connected, wireless nodes, while implementing self-organization network techniques to optimize the overall network performance.
The attached devices are typically managed via a cloud service (device cloud).
A extremely stripped-down version of the \gls{iiot} edge gateway could be wireless sensor/actor devices themselves, which build up a mesh network among each other.

\subsubsection{\glsentryshort{sdn} Controller}\label{sec_sdn_ctrl}
The \gls{sdn} controller is the centralized management unit of a \gls{sdn}.
It acts as a strategic controller in the \gls{sdn} network, manages flow-tables on \gls{sdn} switches, which contain the data activities in the network using a \gls{sdn} protocol like OpenFlow \cite{open_networking_foundation_openflow_2013} or Open vSwitch Database Management Protocol \cite{pfaff_open_2013}.
Thus, it can configure the network dynamically such that it can meet the changing needs related to configuration, security and network optimization.
OpenDaylight (\url{http://www.opendaylight.org/}, Java-based, supports OpenFlow and other \glspl{api}) and Ryu (\url{http://osrg.github.io/ryu/}, Python-based, supports OpenFlow) are two typical \gls{sdn} controllers. 

\subsubsection{\glsentryshort{sdn} Switch}\label{sec_sdn_switch}
The \gls{sdn} switch is a device, which receives, sends and forwards data packets in a network, in order to meet specific requirements.
On the one hand, \gls{sdn} can be considered being a pure software solution, which flexibly connects multiple instances on one or multiple cloud server hardware, which we term cloud-\gls{sdn}.
On the other hand dedicated, discrete \gls{sdn} hardware (\gls{sdn} switches and routers) is available, which connect through multiple physical ports.
Both follow the rules in a flow-table, which is managed by the \gls{sdn} switch via \gls{sdn} protocols, like OpenFlow.
Dedicated \gls{sdn} switch hardware is not required, even though many vendors offer specific \gls{sdn} switches to deliver enhanced \gls{sdn} performance.
There are some virtual \gls{sdn}-Switches, e.g. \gls{ovs}, to provide a switching stack for hardware virtualization environments.

\subsubsection{Radio Management}
The radio management entity gathers context information about the radio environment and particularly about the link quality of the wireless links associated with the \gls{ap}.
This information can be used to apply self-x mechanisms to the overall connectivity solution using its \gls{sdn} capabilities.
If a device is connected to the \gls{ap} using two independent WiFi adapters, only one wireless link would be established using one frequency band through one WiFi interface, in a first step. 
The internal \gls{sdn}-switch is configured such that traffic going through this interface is routed towards the cloud. 
The second WiFi interface is used for wireless monitoring. 
Both interfaces send context information about the current wireless status to the radio resource management. 
If then the utilization of a different channel would be beneficial, a second WiFi link is established using another frequency band. 
The \gls{sdn} switch is reconfigured such that data is now sent through the second wireless interface.
Then, the first wireless link can be shut down such the process is \enquote{make before break} and runs seamlessly without packet loss. 
Alternatively, the \gls{sdn} switch could implement multipath techniques such as Multipath TCP or \gls{prp}.



\section{Conclusion}
In this paper, a result of the \enquote{5Gang} project is presented.
Use cases motivated by actual needs of the industry project partners are stated.
The architecture is general in such a way that it offers the possibility to connect devices of all use cases of the \enquote{5Gang} project as well as of \cite{3gpp_study_2017}.
It fits into the existing frameworks, builds upon existing 5G architectures and remains flexible such that it can server future needs of the factory automation.

\section*{Acknowledgment}
This work has been supported by the Federal Ministry of Education and Research of the Federal Republic of Germany (Foerderkennzeichen 16KIS0725K, 5Gang). 
The authors alone are responsible for the content of the paper.

\bibliography{5GangArch}
\bibliographystyle{IEEEtran}


\end{document}